# Chapter 12: DNA methylation markers to assess biological age

Dmitriy I. Podolskiy[1,*], Vadim N. Gladyshev[1,*]

[1] Division of Genetics, Department of Medicine, Brigham and Women's Hospital and Harvard Medical School, Boston, MA 02115, USA

[*]V.N.G. (vgladyshev@rics.bwh.harvard.edu), D.I.P. (dpodolskiy@research.bwh.harvard.edu)

**Summary**

Among the different biomarkers of aging based on omics and clinical data, DNA methylation clocks stand apart providing unmatched accuracy in assessing the biological age of both humans and animal models of aging. Here, we discuss robustness of DNA methylation clocks and bounds on their out-of-sample performance and review computational strategies for development of the clocks.

Discovery of a precise and robust molecular biomarker of age opens an important venue in aging research: the practical possibility to perform truly large-scale screens for dietary, pharmacological and genetic interventions extending lifespan of model animals and, potentially, of humans. Indeed, a critical limitation of any such screen is *time available* to do the experiment and determine its outcome. To be sure whether a given intervention truly extends the lifespan of the organism under consideration, a rule of thumb is to maintain control animals for at least a couple of mortality doubling times (~8 months for mouse) to be able to detect the effect of the intervention on the structure of the resulting survival curve. Thus, for long-lived organisms such as the naked mole rat (or, for that matter, humans with the mortality rate doubling time of ~8 years) the time of the experiment becomes large enough to make any such screening study impractical. Moreover, as interventions not influencing the mean lifespan may nevertheless modulate the maximal lifespan achievable within the cohort, it is ideal to wait for the whole cohort to die out to have access to the complete survival curve. This requirement is even more limiting; for example, the maximal lifespan of C57Bl/6 mice routinely used as a control in lifespan screens is ~3 years, and some interventions (calorie restriction, growth hormone receptor knockout, etc.) can increase it to 4-4.5 years.

While the lack of available time may be somewhat addressed by increasing the cohort size under study, the compensating effect is limited: recall that the survival curves of most organisms used as models to study human aging are largely determined by Gompertz law dictating that all-cause mortality increases exponentially with age, and as such, the corresponding survival distributions feature super-exponentially decaying tails at ages larger than the mean lifespan. Consequently, when one identifies parameters of these survival distributions, the identification error depends on the cohort size sub-logarithmically weakly. Thus, we must conclude that the limitation of time available to perform a lifespan intervention screen and determine its results with statistical significance is of fundamental nature.

If developed, a good experimental biomarker of age would address this limitation, allowing to measure the biological age of an organism subjected to a lifespan-modulating intervention, and compare it to a control obtaining the estimate on effectiveness of the intervention right away. Early attempts to develop such a biomarker were based on modulation of gene expression levels with age [1], change in somatic mutation counts with age [2–5], shortening of telomeres [6], and other approaches. Unfortunately, none of them proved to be sufficiently accurate to be used practically in large scale screens, but the situation finally may have changed with the discovery of DNA methylation clocks in humans [7,8] and other mammals [9–13].

Average DNA methylation across CpG sites present in the human genome or genomes of several animal models is known to systematically change with age [14]. For example, in both human and mouse the wave of increased methylation is strongly pronounced during the early period of organismal development [7–9]; it is then replaced by a relatively slow process of hypomethylation (on average across the genome) proceeding through the remaining lifespan of the organism (Fig. 1). Exploiting these systematic age-dependent processes allows to develop a potentially very precise biomarker of age in both human (with accuracy ~3-5 years in identifying the sample age [8]) and mouse (with reported accuracy as high as ~1 month for some DNA methylation clocks and mouse strains [9]).

As reviews and comparative analyses of already developed DNA methylation clocks are presented in other chapters of this book (see Chapters 11 and 13), we devote this short chapter to the discussion of robustness of methylation clocks, their out-of-sample performance and the computational strategies for robust DNA methylation clock construction. We shall also discuss several possible pitfalls which researchers may encounter in the process of biomarker development; this discussion is largely applicable to any other biomarkers of aging based on a regularized linear or generalized regression of an arbitrary omics dataset to chronological ages of samples – including those constructed using neural networks (see Chapter 18). We assume that the reader is familiar with the biology of the process of DNA methylation, its mechanics, dynamics and significance in the regulation of gene expression [14].

**Building the clock: the meaning of regularization parameters and number of CpG sites vs. the number of samples**

Canonically [8], once a dataset describing an age-dependent modulation of methylation levels on CpG sites across the genome is obtained, the CpG sites contributing to the potential clock are identified by performing elastic net regression of DNA methylation levels to the chronological age of organisms, from which the samples are collected.

Elastic net regression is a regularized regression of a superposition of stochastic variables $X$, in our case methylation levels $X \in [0,1]$ on CpG sites in the genome numerated by the index $j$ with its value getting as high as several million since there are ~30 million CpG sites in the mammalian genome [14]; in practice, $1 < j < n$ reaches values of ~2-4 million for samples obtained by reduced representation bisulfite sequencing (RRBS [15] in what follows) and remains $\lesssim 1$ million for samples assayed with Illumina chips. The regression process seeks to minimize the target function

$$\frac{1}{2N}\sum_{i=1}^{N}(Age_i - \beta_0 - \beta_i X_i)^2 + \lambda T_\alpha(\beta), \quad (1)$$

with respect to the values of the intercept $\beta_0$ and the weights of individual predictors $\beta$; here

$$T_\alpha(\beta) = \alpha||\beta||_{L1} + \frac{1-\alpha}{2}||\beta||_{L2}^2 = \sum_{j=1}^{n}\left(\alpha|\beta_j| + \frac{1-\alpha}{2}\beta_j^2\right), \quad (2)$$

$N$ is a number of observations/samples with chronological ages $Age_i$, $||\beta||_{L1}$ and $||\beta||_{L2}$ are the standard $L_1-$ and $L_2$-norms, respectively.

The purpose of extending the regularization norm (2) beyond a simple $L_1$-norm of the least absolute shrinkage and selection operator (LASSO) regression is to make the target function for the regression strictly convex [16], since the target function of LASSO regression is not guaranteed to possess this property. For the target function (1), (2) there exists a single global minimum, and, if the number of predictors $n$ is not very large, there is hope that the global minimum will be reached within a small number of iteration steps. Also, when the number of samples $N$ is smaller than the number of predictors $n$ (which is a typical situation in any omics-related problem and an especially saturated one in the case of the analysis of DNA methylation data characterized by the numbers of samples counted in hundreds and the numbers of observables in millions), the LASSO regression procedure based on $L_1$-norm typically chooses the number of clock sites $< N$ (number of samples). It is easy to spot-check that virtually all DNA methylation clocks built using elastic net regression satisfy the very same property [7–11], and thus elastic net regression inherits this drawback. As methylation levels on many CpG sites are strongly correlated between each other (this is the case for CpG islands and, more generally, for promoters of gene expression [14]), both LASSO regression and its extension based on elastic net also tend to pick a limited number of CpG sites within the correlated subset and ignore the others.

The meaning of the regularization parameters in the target function (1), (2) is as follows. First, one expects that increasing the relative weight $\alpha$ of the $L_1$-norm makes the global minimum of the target function (1), (2) less pronounced; lower $\alpha$ are therefore preferable. On the other hand, higher $\alpha$ means that vectors of weights $\beta$ with larger $L_2$-norms $\sim |\beta|^2$ are penalized more strongly; this has a potential to decrease the number of CpG sites contributing to the constructed clock and therefore negatively influence its robustness. As for the overall weight $\lambda$ of the regularizing part of the target function, it should be chosen according to the comparative performance of the clocks built with different values of $\lambda$. In a situation of a single global minimum of the target function one hopes that the drawbacks of LASSO regression discussed above are avoided. However, as we discuss below, this raises another problem: the target function will possess a single global minimum even in the case of significant amount of noise present in the analyzed dataset.

**Two illustrative examples. Notes on robustness and accuracy of methylation clocks out-of-sample**

Generally, CpG sites in a mammalian genome are strongly non-equal biologically (some among them are deeply "buried" within the genome and virtually never accessible, while others are almost always accessible and their methylation levels promote or suppress the expression of genes [14]), and elastic net regression of DNA methylomes to chronological age would in principle allow to identify these important sites with methylation levels systematically changing with age. Let us, however, imagine that we are dealing with datasets characterized by very low signal-to-noise ratio (often datasets obtained by RRBS belong to exactly this category, as methylated and unmethylated read counts on individual CpG sites might fluctuate widely from sample to sample). Let us determine the answer, which a standard procedure might give in that case. Namely, we shall perform the following simple experiment: generate a number $N$ of methylome snapshots covering $n$ CpG sites, such as when the methylation level on each site is a random number uniformly distributed in the interval [0,1]; each methylome should also be assigned a "chronological age", and we keep $n \gg N$ generating this dataset. To maximize the number of clock sites (robustness) and simultaneously ensure that the target function possesses a single global minimum, we choose $\alpha = 1/2$ providing the same weight to $L_1$- and $L_2$-measures.

It is easy to check that subjecting the resulting dataset to the procedure of elastic net regression with 20-fold cross-validation results in a rather "precise clock" based on ~200 CpG sites out of 60,000, see Fig. 2, 3. Obviously, performance of this "clock" will be abysmal on a test set (which can be generated in the same fashion as the training set), but what have we really observed here? How was it even possible to construct a seemingly precise clock on a completely random data? The answer is that elastic net regression has very effectively performed noise fitting: when the number of observations is much smaller than the number of predictor variables (CpG sites), one will always find a subset of predictor variables which by a mere chance behave coherently as functions of chronological age, and the regularized regression will necessarily pick this subset to include it in the resulting clock.

While this experiment is illustrative, it does rarely correspond to the situation one faces when building the clock from the real DNA methylation datasets as its precision out-of-sample is negligible. We thus consider now a somewhat more advanced experiment: take the same number of samples $N$, assign them monotonously increasing chronological ages and assume that the methylation levels on every CpG site among $n$ included in the simulated methylomes grow linearly as the functions of chronological age of samples. In addition, we shall also prescribe a uniformly distributed noise with the same amplitude,

$$X_i(Age) = (c \cdot Age + f(Age, i))/|c \cdot Age + f(Age, i)|,$$

to every CpG site in the methylome generated in the same fashion as considered in the previous numerical experiment and renormalize values of the methylation levels such that they remain defined in the interval [0,1].

Subjecting the resulting dataset to the procedure of elastic net regression with $\alpha = 1/2$ and 20-fold cross-validation again reveals a rather precise clock with ~19 days on the training set and ~32 days on the test set of a similar size generated in a similar fashion, Fig. 4, 5. Two important observations are now in order:

(1) There is a noticeable deterioration of clock performance on the test set. Obviously, it takes place because by construction any CpG site within simulated methylomes can be considered as a clock site. What the procedure of elastic net regression does in this case is again partly fitting noise because realizations of the random noise on some among $n$ CpG sites will look like a consistent age-dependent signal, and the fitting procedure will pick them up. On the test set, there will be another combination of CpG sites, different from the constructed clock, characterized by the realization of the random noise looking the closest to a consistent age-dependent signal, etc.

(2) The performance reduction on the test set is really marginal since *any subset* of CpG sites included in the simulated methylomes will generate a clock with a comparable performance out-of-sample. Performance will also be relatively similar for clocks with any large (or small) number of the CpG sites contributing to them, but the clocks with larger numbers of clock sites will be characterized by a more robust performance out-of-sample.

By construction, we did not consider here CpG sites with biological relevance and a robust age-dependent signal, but the take-away from the two examples presented here is that due to the existence of a global minimum of the target function on any training dataset and in a situation when the number of samples in the training set is much smaller than the number of predictor variables, one will often end up with a clock with a relatively low performance out-of-sample, as the elastic net regression will partially fit noise present in the dataset in this case. Let us now obtain an estimate for the upper bound on the clock performance out-of-sample.

**Estimating an upper bound on out-of-sample performance of methylation clocks**

A general upper bound on the out-of-sample performance of a methylation clock can be obtained by performing a standard robustness analysis of the constructed clock. This analysis generally includes adding a very large number of realizations of a random noise to the training dataset and obtaining estimations of the biological age from the resulting dataset. Subsequently, the upper bound on the performance error out-of-sample is obtained by averaging over realizations of the random noise.

In the corresponding theoretical calculation, one considers a worst-case scenario, when the signal encoded in these counts is well below the level of noise ("maximal entropy" estimation). For simplicity, one consider fluctuations of methylated and unmethylated reads count on every clock site as two uncorrelated Gaussian-distributed stochastic variables. Since the methylation level on a CpG site is defined as $M_i =$

$\frac{Reads_{met}}{Reads_{unmet}+Reads_{met}}$, the denominator is also a Gaussian-distributed stochastic quantity with a mean $E(Reads) = E(Reads_{unmet}) + E(Reads_{met})$ and a variance $\sigma^2 = \sigma_{unmet}^2 + \sigma_{met}^2$. Performing the Geary-Hinkley transformation [17] on the stochastic variable $M_i$ one finds that the stochastic variable

$$t \approx \frac{E(Reads)M_i - E(Reads_{met})}{\sqrt{\sigma^2 M_i^2 + \sigma_{met}^2 - 2\rho\sigma\sigma_{met}M_i}} \quad (3)$$

is normally distributed with zero mean and unit variance, and it is thus possible to obtain an approximate form for the distribution function of the methylation levels $M_i$. In particular, the expression for the error in the total weighted average methylation can now be bounded from above as

$$\approx \sum_{j=1}^{n} |\beta_j| E(\delta M_j^2) \lesssim \frac{1}{\sqrt{2}} ||\beta||_{L_1},$$

i.e., by the $L_1$-norm of the vector of clock weights. The bound on the error of the actual perceived methylation age is in turn given by

$$E(\delta Age^2) \lesssim \sqrt{2} f''\left(E\left(\beta_0 + \sum_{j=1}^{n} \beta_j M_j\right)\right) ||\beta||_{L_1}, \quad (4)$$

where the methylation age is related to the weighted average methylation $E(M)$ according to the functional dependence

$$Age = f(E(M)),$$

i.e., the upper bound on the clock performance out of sample is also roughly determined by the convexity of the functional dependence $f(x)$ and the $L_1$-measure of the vector of clock weights. It follows immediately from the expression (4) that the clocks determined by a larger number of clock sites $n$ will have a relative performance error lowered by a factor of $1/\sqrt{n}$ (according to the law of large numbers) at early chronological ages. As a "on the back of the envelope" estimate of the bound on the relative out-of-sample clock performance in this regime one can get as a formula

$$\frac{\sqrt{E(\delta Age^2)}}{E(Age)} \sim \frac{1}{\sqrt{n \cdot E(\beta)}}, \quad (5)$$

where $n$ is the number of CpG sites contributing to the clock and $E(\beta)$ is an average weight of a clock site in the expression for the weighted average methylation. The error $\sqrt{E(\delta Age^2)}$ grows with age monotonously [9–11], and thus the estimates (4), (5) really serve as very rough lower bounds on the performance of the clock out-of-sample in the worst case scenario of a dataset with low signal-to-noise ratio. It is also easy to check that the estimate (4) does not strongly depend on the distribution properties of methylated/unmethylated read counts (since the stochastic variable $M_i$ is always bounded between 0 and 1).

## Conclusion

Based on the analysis presented above, we would like to conclude that to build a DNA methylation clock using relatively noisy DNA methylation datasets such as those obtained by RRBS it might be beneficial to use a canonical LASSO regression, which corresponds to elastic net regression with $\alpha = 1$. LASSO regression does not excessively penalize clocks with larger numbers of CpG sites contributing to them. As such, it would generate a methylation clock with performance out-of-sample comparable to the clocks based on elastic net regression, and a clock based on a larger number of CpG sites will also have a higher robustness when applied to samples with several clock sites not covered. The drawback of LASSO on the other hand is that its target function is not guaranteed to possess a single global minimum.

The upper bound on the out-of-sample performance of elastic net regression clocks obtained here is uncomfortably high. For example, for the 90-site blood clock developed for mouse methylomes one finds that out of sample the error of age identification can go in probability as high as $\sqrt{E(\delta Age^2)} \sim 1.5 \cdot E(Age)$. Again, essentially the only parameter left for controlling this bound is the number of clock sites, and a clock including thousands of sites would lead to lowering this bound by an order of magnitude.

Recalling that both elastic net and LASSO regressions typically converge to a clock with a number of sites lower than the number of samples used to train it, we conclude that the increase in the number of sequenced samples will necessarily lead to the emergence of a new generation of DNA methylation clocks with significantly better performance accuracy and robustness.

**Figure legends**

**Fig. 1.** Average methylation across ~2 million CpG sites in the genome of C57Bl/6 male mice, data from [9]. The wave of increased methylation during development (1st month of life) is replaced by monotonous overall hypomethylation during the adulthood.

**Fig. 2.** Building a clock on a random dataset. We used 585 samples of random DNA methylomes containing 60,000 sites each. Behavior of the deviance error of the clock as a function of regularization parameter $\lambda$. The target function has a global minimum even on a completely random dataset. The green line corresponds to the minimum of the deviance function, and the blue line denotes a clock separated from the minimal deviance clock by 1 standard deviation.

**Fig. 3.** Performance of a clock built on a random dataset. The weighted average methylation is clearly correlated with the "age" of samples. Performance of this clock on a test (random) dataset is negligible.

**Fig. 4.** Building a clock in a situation when every CpG site in the methylome is used as a clock site. Elastic net regression chooses 242 sites among 60,000 which produce the best clock. These sites are however chosen by fitting noise on top of the monotonous age-dependent signal present on every site; a particular realization of the random noise imprinted within the training set creates a perception that behavior of methylation levels on some sites with age is more systematic than the others. Again, the target function of the elastic net regression possesses a single global minimum as a function of optimization parameters.

**Fig. 5.** Performance of the clock from Fig. 4. Shown is the performance on the training set (the standard deviation of the methylation age from the chronological age is ~19 days) and on the test (the same error is ~32 days). The error on a test sample is low because any CpG site among 60,000 covered can be used as a clock site. Similarly, any other clock built using an arbitrary linear combination of covered CpG sites would produce a similar performance out-of-sample.

**Figures**

**Fig. 1**

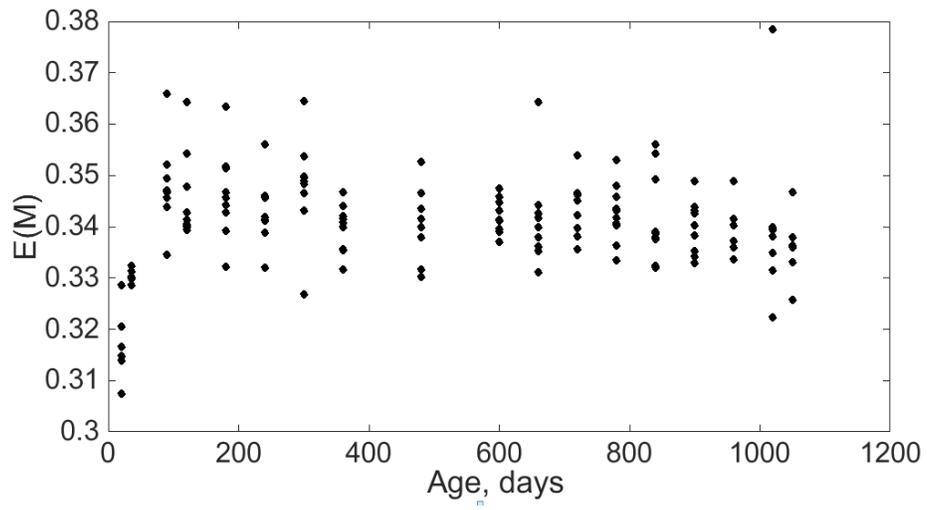

**Fig. 2**

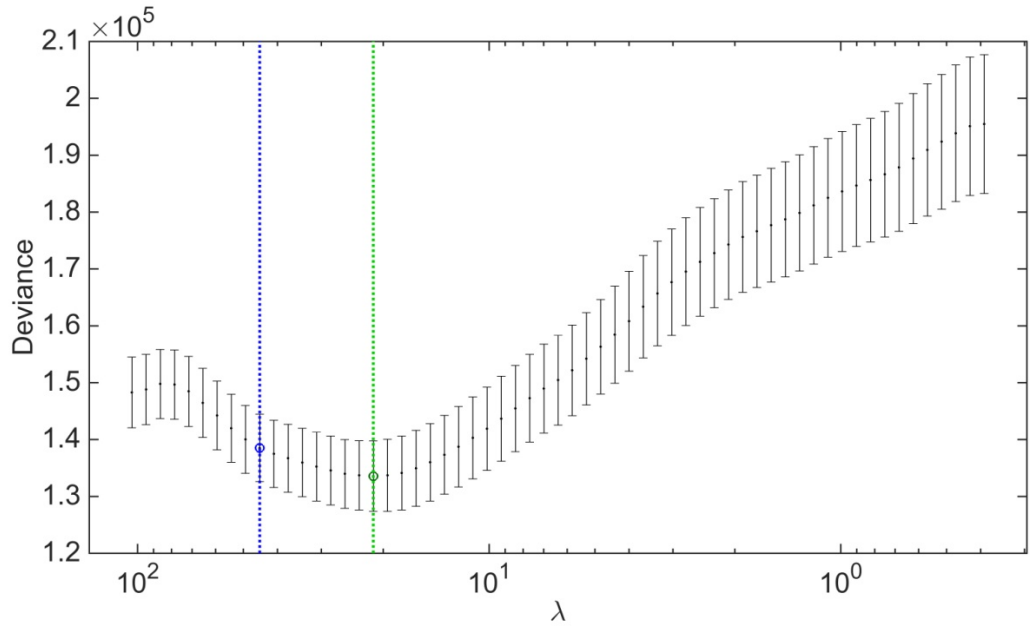

**Fig. 3**

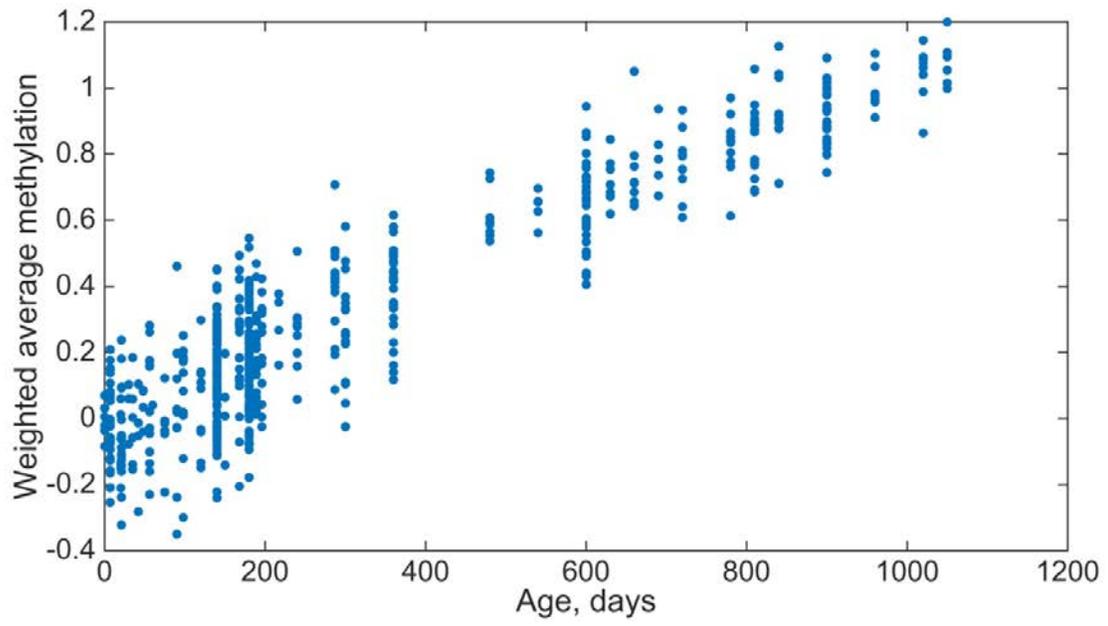

**Fig. 4**

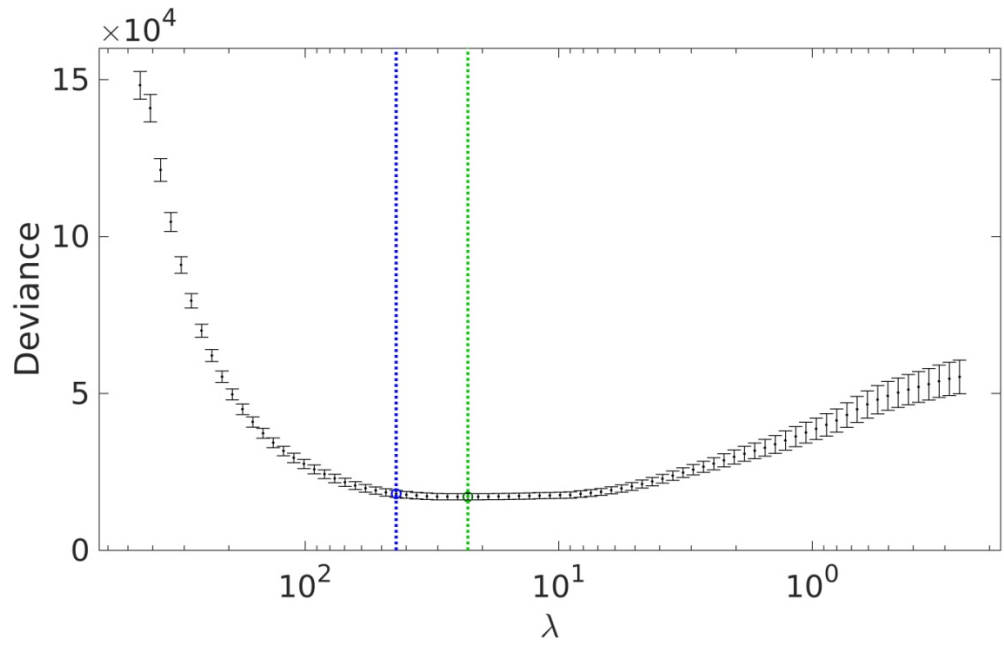

**Fig. 5**

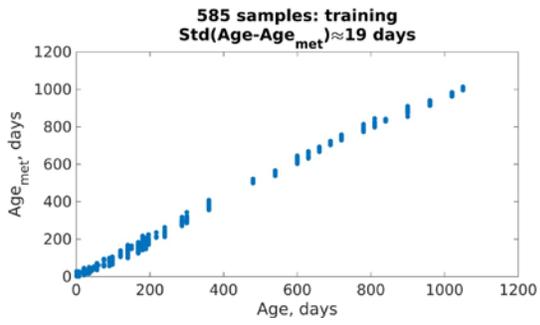 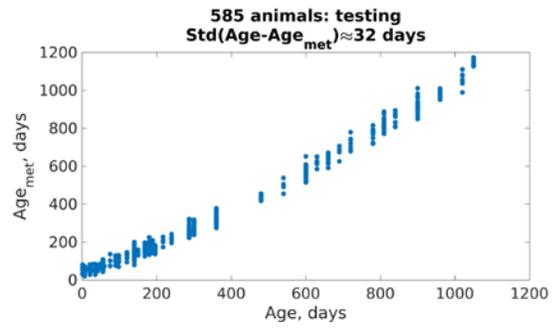